\documentclass[10pt,a4paper,twocolumn]{article}
 
\usepackage[margin=1.8cm, headheight=14.5pt]{geometry}
\usepackage{amsmath,amssymb}
\usepackage{graphicx}
\usepackage{hyperref}
\usepackage{xcolor}
\usepackage{booktabs}
\usepackage{enumitem}
\usepackage{caption}
\usepackage{subcaption}
\usepackage{float}
\usepackage{fancyhdr}
\usepackage{titlesec}
\usepackage{abstract}
\usepackage{tabularx}
\usepackage{verbatim}
\usepackage{microtype}

\hypersetup{
  colorlinks=true,
  linkcolor=blue!70!black,
  citecolor=green!50!black,
  urlcolor=blue!60!black
}
 
\title{%
  {\LARGE\bfseries Code Broker: A Multi-Agent System\\[4pt]
  for Automated Code Quality Assessment}\\[10pt]
}
 
\author{%
  Samer Attrah \\
  Independent researcher \\
  California, USA \\
  samiratra95@gmail.com \\
}
 
\date{May 2026}
 
\begin{document}
 
\twocolumn[
  \begin{@twocolumnfalse}
    \maketitle
    \thispagestyle{fancy}
    \begin{abstract}
    \noindent
    We present \textbf{Code Broker}, a multi-agent system built on Google's Agent Development Kit (ADK) that analyses Python source code---from individual files, local directory trees, or remote GitHub repositories---and generates structured, actionable quality-assessment reports. The system realises a hierarchical five-agent architecture in which a root orchestrator coordinates a sequential pipeline agent that, in turn, dispatches three specialised agents concurrently---a Correctness Assessor, a Style Assessor, and a Description Generator---before synthesising their findings through an Improvement Recommender. Reports quantify four quality dimensions---correctness, security, style, and maintainability---on a normalised scale and are rendered in both Markdown and HTML for integration into diverse developer workflows. Code Broker fuses LLM-based semantic reasoning with deterministic static-analysis signals from Pylint, employs asynchronous execution with exponential-backoff retry logic to improve robustness under transient API failures, and explores lightweight session memory for retaining and querying prior assessment context across runs. We frame this paper as a technical report on system design, prompt engineering, and tool orchestration, and present a preliminary qualitative evaluation on representative Python codebases of varying scale. The results indicate that parallel specialised agents produce readable, developer-oriented feedback that complements traditional linting, while also foregrounding current limitations in evaluation depth, security tooling, large-repository handling, and the exclusive reliance on in-memory persistence. All code and reproducibility materials are publicly available~\cite{codebrokerdev}.
    \end{abstract}
    \vspace{0.5cm}
  \end{@twocolumnfalse}
]
 
\section{Introduction}
 
Software quality assessment is an indispensable yet resource-intensive activity in the software development lifecycle. Industry data consistently show that remediation costs escalate dramatically with latency: a defect caught after release is estimated to cost 25--30$\times$ more to resolve than one identified during the review phase. Manual code review, although effective, is susceptible to reviewer fatigue, cognitive overload, and throughput bottlenecks---particularly in organisations where senior engineers may dedicate upward of half their working hours to reviewing others' contributions rather than authoring new functionality. Deterministic static-analysis tools such as Pylint, Flake8, and SonarQube partially alleviate this burden by surfacing syntactic violations and common anti-patterns; however, they operate at the surface level and seldom deliver the nuanced, context-sensitive feedback that a skilled human reviewer would provide.
 
The emergence of large language model (LLM)-powered agents introduces a qualitatively different paradigm: autonomous systems capable of \emph{comprehending} code semantics, reasoning about logical correctness, detecting security anti-patterns, and articulating domain-specific improvement recommendations---all expressed in natural language. Code Broker explores this paradigm by constructing a multi-agent system that synergises the precision of traditional static analysis with the contextual reasoning capabilities of contemporary LLM agents.
 
Code Broker was developed as part of the \emph{Google/Kaggle 5-Day AI Agents Intensive} capstone project~\cite{google2026intensive}. The current implementation targets Python codebases and accommodates three input modalities: individual files, local directory trees, and remote GitHub repositories.
 
\subsection{Motivation}
 
The core motivation behind Code Broker is to democratise access to high-quality code review. Junior developers, solo practitioners, and small teams frequently lack access to experienced reviewers who can provide rapid, quantitative assessment with calibrated quality metrics. Code Broker addresses this gap by delivering fast, structured feedback that transcends line-level linting to encompass architectural design, logical correctness, security posture, and long-term maintainability. Equally important, the system is designed to assist software engineers in comprehending and assessing unfamiliar codebases \emph{without} the need to exhaustively explore every file---a capability that becomes especially valuable for large repositories and sprawling multi-module projects, where manual familiarisation can consume hours or days before any meaningful review even begins. Moreover, a multi-agent architecture enables \emph{concurrent} specialised analysis rather than funnelling all reasoning through a single monolithic prompt, thereby reducing latency and promoting independence of assessment perspectives.
 
\subsection{Contributions}
 
The primary contributions of this work are:
\begin{enumerate}
  \item A working five-agent hierarchical architecture for code quality assessment, implemented with Google ADK.
  \item A parallel assessment strategy decoupling correctness, style, and description analysis for speed and modularity.
  \item Integration of Pylint as an ADK tool, grounding LLM outputs with concrete static-analysis evidence.
  \item A preliminary qualitative evaluation on representative Python codebases, framed as an initial system study rather than a full benchmark.
  \item Open-source release of all code, notebooks, and documentation~\cite{codebrokerdev}.
\end{enumerate}
 
\section{Related Work}
 
The development of Code Broker is situated at the intersection of multi-agent systems, automated software engineering, and large language model (LLM) orchestration. This section reviews the theoretical and practical foundations that underpin our proposed architecture. We begin by defining AI agents and the multi-agent paradigms that enable complex task decomposition. Next, we describe the Google Agent Development Kit (ADK), the framework utilized for constructing our hierarchical agent pipeline. We then survey contemporary approaches to automated code review, highlighting the shift from deterministic static analysis to context-aware LLM agents. Finally, we discuss emerging research in human-agent collaboration and normative coordination, which provides the sociotechnical context for deploying such systems in real-world development environments.

\subsection{AI Agents and Multi-Agent Systems}
 
An AI agent is an autonomous system that perceives its environment, reasons about goals, selects actions, and executes those actions through tool invocation~\cite{yao2023react,weng2023agent}. The ReAct paradigm~\cite{yao2023react} forms a conceptual basis for many contemporary LLM agents by interleaving reasoning traces with action steps. Multi-agent systems (MAS) extend this foundation by decomposing complex tasks across specialised agents coordinated by an orchestrator~\cite{cai2025designing}. Recent work on \emph{externalization} in LLM agents~\cite{zhou2026externalization} argues that practical agent progress increasingly depends on externalizing cognitive burdens---state into memory, expertise into skills, and structure into protocols---rather than relying solely on stronger models. In software engineering settings, this decomposition is particularly attractive because code-review tasks naturally partition into evidence gathering, analysis, summarisation, and recommendation stages.
 
\subsection{Google Agent Development Kit (ADK)}
 
Google's ADK is an open-source, model-agnostic framework---first introduced at Google Cloud NEXT 2025---designed to streamline the end-to-end construction, orchestration, evaluation, and deployment of production-grade AI agents and multi-agent systems. ADK employs an event-driven runtime that manages LLM invocations, tool callbacks, and state persistence, and natively supports the Agent-to-Agent (A2A) protocol for secure inter-agent collaboration. Key ADK abstractions leveraged by Code Broker include:
 
\begin{itemize}
  \item \textbf{LLMAgent}: A language-model-backed agent that accepts a system prompt, a set of tools, and sub-agents.
  \item \textbf{SequentialAgent}: Chains agents in a fixed order, passing state between steps.
  \item \textbf{ParallelAgent}: Dispatches multiple child agents concurrently and merges their outputs.
  \item \textbf{Runner}: Manages the execution lifecycle, session state, and event loop.
  \item \textbf{AgentTool}: Wraps an agent as a callable tool, enabling agent-of-agents patterns.
  \item \textbf{Memory Service}: Provides session persistence and semantic retrieval via \texttt{InMemoryMemoryService}, enabling context preservation across assessments and query-based memory search.
\end{itemize}
 
\subsection{Automated Code Review}
 
Prior work on automated code review spans three broad and increasingly overlapping categories. \emph{Static analysis tools}~\cite{vassallo2019developers} such as Pylint and SonarQube provide deterministic warnings and style checks with high precision but limited semantic interpretation. \emph{Learning-based approaches}~\cite{lu2021codexglue} support defect prediction and code understanding, yet often depend on benchmark-specific supervision and struggle to generalise across domains. \emph{LLM-based agents}~\cite{hong2023metagpt,qian2023communicative} leverage planning and tool use to generate contextual review feedback, marking a qualitative shift toward natural-language code understanding. Recent systems such as \textit{HyperAgent}~\cite{phan2024hyperagent} and \textit{Agyn}~\cite{benkovich2026agyn} frame software engineering as a collaborative, team-oriented process, while survey work highlights both the promise and reliability challenges of agentic software engineering pipelines~\cite{tang2026challenges,he2024literature}. Notably, the \textit{c-CRAB} benchmark~\cite{zhang2026crab} provides the first systematic evaluation framework for code-review agents, finding that state-of-the-art commercial and open-source review agents collectively address only approximately 40\% of benchmark tasks---underscoring substantial room for improvement. On the retrieval-augmented side, \textit{LAURA}~\cite{zhang2025laura} demonstrates that enriching LLM reviewers with retrieved review exemplars and code-change context can yield correct or helpful comments in over 40\% of cases.
 
Code Broker operates at the intersection of these categories: it employs Gemini for semantic reasoning, Pylint for deterministic evidence, and a hierarchical orchestration pattern to decouple description, correctness, style, and recommendation tasks. Relative to prior work, the objective is not to advance a new theory of automated review but to document a concrete, reproducible code-assessment workflow built with ADK and to examine its practical trade-offs in a capstone setting.

\subsection{Human-Agent Collaboration and Norms}

As agents assume greater autonomy, the strategic allocation of responsibilities between humans and AI agents becomes a first-order design concern. Ronanki~\cite{ronanki2025trust} proposes principles for trustworthy human-agent collaboration, while Dam et al.~\cite{dam2025normative} investigate normative coordination in human-AI engineering teams. Complementarily, Goldman et al.~\cite{goldman2025resolve} find that readability, bug, and maintainability-oriented review comments exhibit the highest developer resolution rates, suggesting that LLM-generated and human reviews are \emph{complementary} rather than substitutive---an insight that directly informs Code Broker's role as an assistive layer. These works provide essential design context, although Code Broker itself remains a developer-assistance tool in which human interpretation of the final report is explicitly required.
 
\section{Methodology}
 
The design of a multi-agent system (MAS) requires navigating a complex landscape of task decomposition, framework affordances, and model performance characteristics. In the development of Code Broker, the orchestration logic---including the specific agent hierarchy and the selection of integrated tools---was engineered to align with the architectural primitives of the Google Agent Development Kit (ADK) and the high-capacity reasoning afforded by the Gemini model family. The resulting configuration balances deep semantic analysis with operational efficiency and bounded execution latency. This section details the system's architecture, individual agent roles, and the integrated methodology employed for code quality evaluation.

\subsection{Overview}
 
\begin{figure*}[t]
\centering
\includegraphics[width=0.86\textwidth]{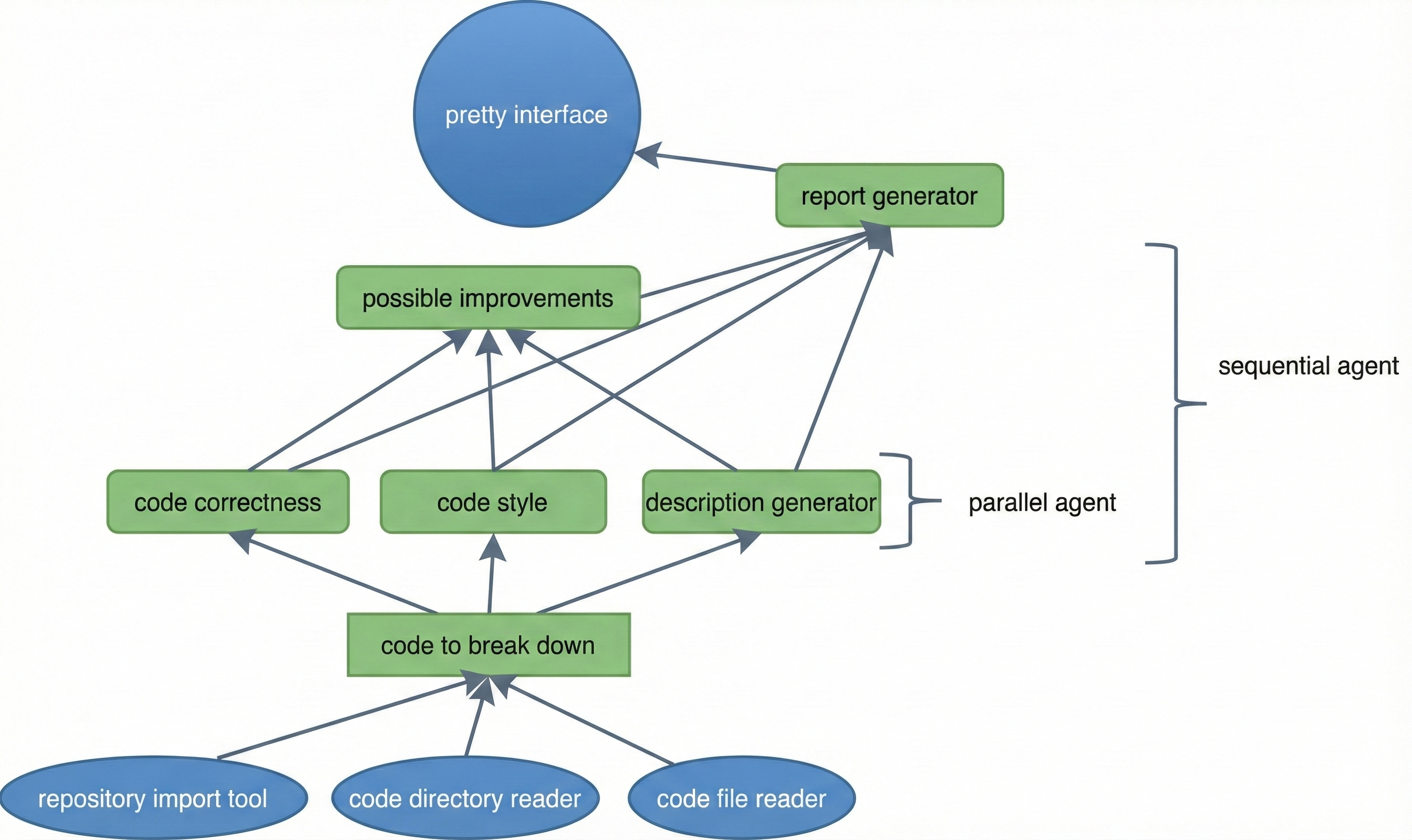}
\caption{Hierarchical five-agent architecture of Code Broker. The orchestrator coordinates a sequential pipeline, which fans-out to three parallel assessors before the Improvement Recommender synthesises a final report.}
\label{fig:architecture}
\end{figure*}

Code Broker is organised as a \emph{hierarchical multi-agent system} with five distinct agents arranged in two layers (Figure~\ref{fig:architecture}). The orchestrator coordinates a pipeline which fans out to parallel specialists before a synthesiser merges results. The design goal is modularity: each agent has a narrow role, an explicit output contract, and limited responsibility for a single review dimension.
 
\subsection{Agent Descriptions}
 
\paragraph{1. Report Generator (Orchestrator).}
The root agent of the system. It accepts the user's input (file path, directory path, or GitHub URL), invokes the Sequential Pipeline Agent as a sub-agent tool, and formats the final consolidated report in Markdown and HTML. Its system prompt instructs it to maintain a professional tone and to surface actionable, developer-friendly language.
 
\paragraph{2. Sequential Pipeline Agent.}
A \texttt{SequentialAgent} that manages the overall assessment workflow. It first invokes the Parallel Assessment Agent and, once all parallel outputs are available, hands results to the Improvement Recommender. It also handles pre-processing: 1) reading file contents, 2) cloning or fetching GitHub repositories, and 3) chunking large files if necessary.
 
\paragraph{3. Parallel Assessment Agent.}
A \texttt{ParallelAgent} that dispatches three child agents simultaneously, reducing total latency. It merges their independent outputs into a structured intermediate representation consumed by the Improvement Recommender.
 
\paragraph{4a. Correctness Assessor.}
Analyses the logical and functional correctness of the code. It checks for algorithmic errors, off-by-one mistakes, improper error handling, and potential runtime exceptions. It also runs Pylint via an ADK tool and incorporates the output into its reasoning. In the current system, security-related observations are also surfaced here when they are inferable from code structure or lint findings.
 
\paragraph{4b. Style Assessor.}
Evaluates PEP~8 compliance, naming conventions, code organisation, and readability. It examines documentation coverage (docstrings, inline comments), complexity metrics, and adherence to project conventions where detectable.
 
\paragraph{4c. Description Generator.}
Produces a concise natural-language summary of what the code does—its purpose, main components, and architectural patterns. This grounds the other assessors and provides context for the final report.
 
\paragraph{5. Improvement Recommender.}
Synthesises all upstream outputs into a prioritised list of improvement recommendations. It scores the codebase on four dimensions (Table~\ref{tab:scoring}) and produces the final report.
 
\subsection{Scoring Dimensions}
 
\begin{table}[ht]
\centering
\caption{Code Broker scoring dimensions (0--10 scale).}
\label{tab:scoring}
\footnotesize
\begin{tabularx}{\columnwidth}{>{\raggedright\arraybackslash}p{1.9cm}X}
\toprule
\textbf{Dimension} & \textbf{Description} \\
\midrule
Correctness & Logical and functional accuracy; presence of bugs, runtime errors, or incorrect algorithmic behaviour. \\
Security & Heuristic vulnerability assessment: possible injection risks, unsafe deserialization, hard-coded secrets, and insecure dependency usage inferred from code review. The current implementation does not yet include a dedicated security scanner. \\
Style & PEP~8 compliance, naming conventions, code readability, documentation quality. \\
Maintainability & Modularity, coupling, cohesion, test coverage indicators, complexity metrics. \\
\bottomrule
\end{tabularx}
\end{table}
 
\subsection{Implementation Details and Tool Integration}

The current prototype was developed and exercised as a notebook-driven system around Google ADK components, with the report generator acting as the top-level entry point. Prompts were manually iterated during the capstone development cycle and tuned for structured output, evidence citation, and concise recommendation synthesis. The implementation should therefore be read as a reproducible system report, but not yet as a controlled study of prompt variants or orchestration strategies.

In parallel with the notebook prototype, a separate packaging effort has been started~\cite{codebrokerpkg} to migrate Code Broker into a reusable Python distribution with a \texttt{src/code\_broker} layout, a \texttt{pyproject.toml} build definition, and a \texttt{code-broker} command-line entry point. This packaging work is relevant to the research trajectory because it pushes the system from a course artifact toward a deployable developer tool, with clearer module boundaries for agents, tools, configuration, reporting, and runtime orchestration.
 
\subsection{Technology Stack}
 
Code Broker is implemented in Python 3.14 and relies on the following key dependencies:
 
\begin{itemize}
  \item \textbf{google-adk}: Multi-agent orchestration framework.
  \item \textbf{google-generativeai}: Gemini model API access.
  \item \textbf{PyGithub}: GitHub repository fetching and file enumeration.
  \item \textbf{pylint}: Static Python code analysis, invoked as a subprocess tool.
  \item \textbf{Jupyter / IPython}: Interactive execution environment via \texttt{notebooks/code\_broker.ipynb}.
  \item \textbf{markdown2}: Markdown-to-HTML rendering for report output.
\end{itemize}

\subsection{Reproducibility Notes}

The implementation uses Gemini-family models accessed through the Google generative AI stack, but the exact serving model can be configured externally in the runtime environment. For this reason, outputs should be considered \emph{configuration-dependent}. In the experiments reported here, we focus on workflow behaviour and report characteristics rather than strict model-to-model comparison. Future versions of the paper should pin the exact model identifier, decoding settings, and prompt templates in an appendix or artifact bundle.

\subsection{Session Memory and Retrieval}

The notebook prototype also includes an initial memory workflow built on ADK's \texttt{InMemoryMemoryService}. After a report-generation run completes, the active session can be written into memory via \texttt{add\_session\_to\_memory}, and subsequent queries can retrieve prior report context through \texttt{search\_memory}. In practical terms, this provides a lightweight mechanism for asking questions such as what previous reports said about a repository or codebase without rerunning the full analysis pipeline.

At present, this memory mechanism remains transient and notebook-scoped: it is useful for demonstrating conversational continuity and report recall, but it is not yet a persistent longitudinal store. This is an important distinction for the research framing. The current implementation shows that memory can enrich post-analysis interaction, while a more mature version would need durable storage, repository-level indexing, and stronger controls over what historical findings are retained and resurfaced.
 
\subsection{Input Handling}
 
The system supports three input modalities:
 
\begin{enumerate}
  \item \textbf{Single file}: The file is read from the local filesystem, chunked if it exceeds the model's context window, and passed directly to the pipeline.
  \item \textbf{Directory}: All Python (\texttt{.py}) files under the target directory are enumerated recursively, and each is assessed individually before the results are aggregated.
  \item \textbf{GitHub repository URL}: The PyGithub library fetches repository metadata and file contents via the GitHub API, respecting rate limits and authentication via an optional \texttt{GITHUB\_TOKEN} environment variable.
\end{enumerate}

For directories and repositories, file-level outputs are aggregated into a repository-level summary by the downstream recommendation stage. When a file is too large for direct inclusion in the prompt context, it is chunked before analysis. This chunking strategy improves coverage, but it can also reduce cross-file and long-range reasoning quality; we therefore treat current large-repository handling as a practical compromise rather than a solved problem.
 
\subsection{Pylint Tool Integration}
 
The Pylint integration is wrapped as an ADK \texttt{FunctionTool}. The tool accepts a code string, writes it to a temporary file, runs Pylint as a subprocess, and returns the structured JSON output. The Correctness Assessor invokes this tool and incorporates the linting findings into its analysis.
 
\begin{figure}[ht]
\captionsetup{type=figure}
\caption{Simplified Pylint ADK tool wrapper.}
\label{lst:pylint}
\footnotesize
\setlength{\tabcolsep}{4pt}
\begin{tabularx}{\columnwidth}{>{\ttfamily\raggedright\arraybackslash}X}
\toprule
import subprocess, tempfile, json \\

def run\_pylint(code: str) -> dict: \\
\quad """Run Pylint on the provided Python source\\
\quad and return results.""" \\
\quad with tempfile.NamedTemporaryFile( \\
\qquad suffix=".py", mode="w", delete=False \\
\quad ) as f: \\
\qquad f.write(code) \\
\qquad tmp\_path = f.name \\
\quad result = subprocess.run( \\
\qquad ["pylint", tmp\_path, "--output-format=json"], \\
\qquad capture\_output=True, text=True \\
\quad ) \\
\quad try: \\
\qquad return json.loads(result.stdout) \\
\quad except json.JSONDecodeError: \\
\qquad return \{"error": result.stdout\} \\
\bottomrule
\end{tabularx}
\end{figure}
 
\subsection{Asynchronous Processing and Retry Logic}
 
All agent invocations are performed asynchronously using Python's \texttt{asyncio}. The ParallelAgent dispatches three coroutines concurrently, and results are gathered with \texttt{asyncio.gather}. A lightweight exponential-backoff retry decorator wraps each agent call to handle transient API errors (rate limits, timeouts). The retry policy uses a maximum of three attempts with jitter. This improves robustness in practice, but it also means end-to-end latency depends on external API behaviour and should not be interpreted as a stable benchmark result.
 
\subsection{Report Generation}
 
The final report is structured as follows:
 
\begin{enumerate}
  \item \textbf{Executive Summary}: One-paragraph description of the codebase generated by the Description Generator.
  \item \textbf{Scores Table}: Four-dimension score table with a brief rationale per dimension.
  \item \textbf{Correctness Analysis}: Detailed findings from the Correctness Assessor, including Pylint output.
  \item \textbf{Style Analysis}: Findings from the Style Assessor.
  \item \textbf{Improvement Recommendations}: Numbered, prioritised list of actions from the Improvement Recommender.
  \item \textbf{Conclusion}: Overall assessment and suggested next steps.
\end{enumerate}
 
Reports are rendered in both Markdown (for downstream toolchain integration) and HTML (for direct browser viewing), with syntax-highlighted code snippets where relevant.
 
\section{Results}
 
\subsection{Evaluation Setup}

We evaluate Code Broker as a \emph{preliminary system study} rather than a full benchmark. The goal of this section is to assess whether the agent pipeline produces coherent, actionable reports across representative Python inputs. Crucially, Code Broker is not only intended for assessing and evaluating code quality, but also for summarizing and facilitating the understanding of code structures and logic, particularly for unfamiliar codebases. Therefore, the evaluation also considers the clarity and accuracy of the generated summaries and their effectiveness in aiding developer comprehension.

The evaluation used a small set of representative Python codebases drawn from three categories:
 
\begin{itemize}
  \item \textbf{Toy scripts}: Small, self-contained utility scripts (50–200 lines).
  \item \textbf{Medium projects}: Open-source utilities with multiple modules (500–2000 lines).
  \item \textbf{GitHub repositories}: Public repositories fetched via the GitHub API.
\end{itemize}

Each case was reviewed manually using four questions: (1) whether the generated description matched the apparent purpose of the code, (2) whether the correctness findings were specific and evidence-backed, (3) whether recommendations were actionable for a developer, and (4) whether the overall report was readable as a stand-alone artifact. We did not use ground-truth bug labels, inter-rater agreement, or a formal baseline in this version of the study; those remain future work.

\subsection{Observed Outcomes}

Across the cases examined, the system generally produced well-structured reports with readable sectioning and concrete next steps. The Description Generator was especially useful for unfamiliar repositories because it provided orientation before the detailed assessment sections. The Correctness Assessor's use of Pylint output improved traceability by anchoring at least part of the review in line-level evidence instead of free-form model claims.

The most consistent weaknesses appeared on larger repositories and in the security dimension. For multi-file projects, chunking and file-level aggregation sometimes reduced architectural coherence. For security, the absence of a dedicated scanner meant that the security score was best interpreted as a heuristic review signal rather than as a comprehensive vulnerability assessment.

\subsection{Report Quality Dimensions}
 
Table~\ref{tab:eval} summarises qualitative observations across evaluation cases.
 
\begin{table}[ht]
\centering
\caption{Preliminary qualitative evaluation summary across representative codebases.}
\label{tab:eval}
\footnotesize
\setlength{\tabcolsep}{3pt}
\begin{tabularx}{\columnwidth}{>{\raggedright\arraybackslash}p{2.15cm}>{\centering\arraybackslash}p{1.25cm}>{\centering\arraybackslash}p{1.45cm}>{\centering\arraybackslash}p{1.35cm}}
\toprule
\textbf{Quality Aspect} & \textbf{Toy Scripts} & \textbf{Medium Projects} & \textbf{GitHub Repos} \\
\midrule
Description Accuracy       & High    & High    & Medium \\
Correctness Coverage       & High    & Medium  & Medium \\
Style Feedback Relevance   & High    & High    & High \\
Recommendation Actionability & High  & High    & Medium \\
Report Readability         & High    & High    & High \\
\bottomrule
\end{tabularx}
\end{table}
 
\subsection{Threats to Validity and Limitations}

The present evaluation and system have the following limitations:
 
\begin{enumerate}
  \item \textbf{Context window constraints}: Very large codebases exceed Gemini's context window, requiring chunking strategies that may fragment logical structure.
  \item \textbf{Limited evaluation protocol}: The study is qualitative, small-scale, and author-assessed. It does not yet include labelled benchmarks, blind human evaluation, or statistical comparison against baselines.
  \item \textbf{Ephemeral memory}: The notebook demonstrates session memory and memory search, but the current setup uses in-memory services only. Memory is therefore not durable across deployments and does not yet provide repository-scale historical tracking.
  \item \textbf{Python-centric}: Pylint integration is Python-specific; extending to other languages requires additional tool wrappers.
  \item \textbf{Weak security grounding}: The system reports a security score, but current security analysis is heuristic and not yet backed by a dedicated scanner such as Bandit.
  \item \textbf{No execution}: The system performs static analysis only; dynamic bugs (e.g., race conditions) may be missed. Future iterations could incorporate formal model checking~\cite{ferrando2024vitamin} or runtime verification layers~\cite{engelmann2022rv} to monitor execution events and ensure normative compliance~\cite{dam2025normative}.
  \item \textbf{LLM hallucinations}: Like all LLM-based systems, the assessors may occasionally produce plausible-sounding but incorrect findings, particularly for domain-specific logic.
  \item \textbf{Rate limits}: GitHub API and Gemini API rate limits can slow analysis of repositories with many files.
\end{enumerate}
 
\section{Discussion}

This section provides a critical analysis of the system's observed behaviour, focusing on the efficacy of the agentic orchestration and the quality of the generated assessments. We examine the impact of prompt engineering on agent output, the resolution of common coordination challenges, and the practical value of the resulting reports for developer comprehension. We further contextualise these findings within the broader trajectory of software engineering, where AI-driven code generation is reshaping development workflows~\cite{kohl2026abundant} and the discipline itself is being redefined around human intent articulation, architectural control, and systematic verification rather than code construction alone.

\subsection{Agent Prompt Engineering}
 
Effective prompt design was critical to achieving high-quality agent outputs. Key principles applied:
 
\paragraph{Role specification.} Each agent's system prompt begins with a concise role statement (e.g., \textit{"You are an expert Python code correctness assessor..."}), setting persona and expertise level.
 
\paragraph{Output format constraints.} Agents are instructed to return structured output with explicit section headers (e.g., \texttt{\#\# Findings}, \texttt{\#\# Score}) to facilitate downstream parsing by the Improvement Recommender.
 
\paragraph{Evidence grounding.} The Correctness Assessor is explicitly instructed to \emph{cite line numbers and Pylint message codes} when reporting issues, reducing vague or unsupported claims.
 
\paragraph{Synthesis instructions.} The Improvement Recommender's prompt instructs it to \emph{de-duplicate} findings from the three parallel agents, \emph{rank} recommendations by severity and impact, and produce at most ten concrete action items.
 
\paragraph{Tone calibration.} All agents are instructed to use constructive, professional language suitable for a developer audience, avoiding condescension or excessive praise.
 
\subsection{Lessons Learned}
 
This project was developed over an intensive five-day period as part of the Google/Kaggle AI Agents Intensive. Key lessons learned:
 
\begin{enumerate}
  \item \textbf{Parallelism improves workflow responsiveness}: Running the three assessors in parallel reduced waiting time in exploratory runs and yielded more independent perspectives, as each agent focused on its specialisation without being influenced by the others. We do not yet report a formal latency benchmark.
 
  \item \textbf{Tool grounding reduces hallucinations}: Incorporating deterministic tools (Pylint) dramatically improved the factual accuracy of the Correctness Assessor. Pure LLM reasoning without tool grounding sometimes missed clear Pylint errors.
 
  \item \textbf{Hierarchical orchestration scales}: The two-layer architecture (Orchestrator $\rightarrow$ Pipeline $\rightarrow$ Parallel Agents) cleanly separated concerns and made it easy to add new assessors without modifying the top-level logic.
 
  \item \textbf{Context management is critical}: Handling large repositories required careful chunking and summarisation strategies to avoid context overflow. Future work will explore retrieval-augmented approaches.
 
  \item \textbf{Prompt iteration is non-trivial}: Agent quality is highly sensitive to prompt wording. Future work will leverage online prompt optimization frameworks like \textit{HiveMind}~\cite{xia2025hivemind} for contribution-guided refinement.
\end{enumerate}

\subsection{Robustness and Reliability}

Ensuring reliability in production-like environments remains a substantive challenge for LLM-based multi-agent systems. We propose adopting chaos engineering methodologies~\cite{owotogbe2025chaos} to proactively surface failure modes such as hallucinated findings, inter-agent communication breakdowns, and cascading retry exhaustion. Furthermore, incorporating dedicated security-analysis agents can mitigate adversarial risks including code injection and prompt poisoning attacks~\cite{bowers2025injection}. As the volume of AI-generated code increases and developer trust in automated outputs faces growing scrutiny, grounding agent claims in verifiable tool evidence---as Code Broker does with Pylint---becomes an essential reliability safeguard.
 
\subsection{Future Directions}
 
Several directions for future improvement are identified:
 
\begin{itemize}
  \item \textbf{Pip package maturation}: Continue the migration from notebook-centric execution to a fully installable \texttt{pip} package with a stable CLI, pinned dependencies, artifact versioning, and test coverage for packaged modules. This would make the system easier to reproduce, evaluate, and integrate into developer workflows.
  \item \textbf{Multi-language support}: Adding language-specific static analysis tools (e.g., ESLint for JavaScript, Checkstyle for Java) to extend beyond Python.
  \item \textbf{Incremental analysis}: Integrating with Git diff APIs to provide assessment of only changed files in pull requests, enabling CI/CD integration.
  \item \textbf{Memory and sessions}: Extending the current notebook-level memory flow into persistent assessment history per repository using ADK session and memory services, enabling longitudinal tracking of code quality, retrieval of earlier reports, and multi-turn follow-up analysis grounded in prior runs.
  \item \textbf{Evaluation benchmark}: Constructing a labelled benchmark of code samples with ground-truth quality annotations for quantitative evaluation.
  \item \textbf{Interactive review mode}: Adding a conversational mode in which developers can ask follow-up questions about specific findings.
  \item \textbf{Security scanner integration}: Incorporating dedicated security scanners (e.g., Bandit for Python) alongside Pylint.
  \item \textbf{Commercial Viability}: Exploring monetization models for Code Broker, such as a premium SaaS tier offering advanced security deep-scans, priority execution infrastructure, and enterprise-grade repository indexing to support sustainable development and professional use-cases.
\end{itemize}
 
\section{Conclusion}
 
Code Broker demonstrates that a hierarchical multi-agent architecture---realised through Google's ADK orchestration primitives---can produce structured, developer-oriented code quality assessments that meaningfully complement traditional static analysis. By parallelising specialised agents, integrating deterministic tools for evidence grounding, and leveraging asynchronous execution with retry logic, the system provides a practical and extensible workflow for exploratory code review in Python projects. As software engineering increasingly shifts from manual code construction toward orchestration and verification of AI-generated artefacts~\cite{kohl2026abundant}, tools like Code Broker occupy a growing niche: bridging the gap between fast but shallow linting and slow but nuanced human review.

The capstone context provided a productive setting for rapidly iterating on agent design, prompt structure, and tool integration under realistic constraints. The next steps for this work are twofold: first, to strengthen the empirical foundation with pinned model configurations, formal baselines such as c-CRAB~\cite{zhang2026crab}, dedicated security tooling, and a benchmark with labelled review outcomes; and second, to complete the transition to a distributable Python package so the system can be rigorously evaluated and adopted outside the notebook environment.
 
\section*{Acknowledgements}
 
The author thanks Google and Kaggle for organising the 5-Day AI Agents Intensive course and capstone competition. Thanks to the Google ADK team for the open-source framework. This project is released under the Apache License 2.0.
 
\bibliographystyle{plain}

\end{document}